\documentclass[
    aps,
    pra,
    reprint,
    showkeys
]{revtex4-2}
\usepackage{amsfonts}
\usepackage{caption}   % ← 加這個
\usepackage{amsmath}
\usepackage{amssymb}
\usepackage{orcidlink}     % 顯
\usepackage{graphicx}%
\usepackage{braket}
\providecommand{\U}[1]{\protect\rule{.1in}{.1in}}
\usepackage{algpseudocode}
\algrenewcommand\algorithmicrequire{\textbf{Input:}}
\algrenewcommand\algorithmicensure{\textbf{Output:}}
\newcounter{algorithm}
\renewcommand{\thealgorithm}{\arabic{algorithm}}

\newenvironment{algorithm}[1][t]
{%
    \begin{figure}[#1]
    \refstepcounter{algorithm}
    \hrule
    \vspace{4pt}
}
{%
    \vspace{4pt}
    \hrule
    \end{figure}
}

\newcommand{\algorithmcaption}[1]{%
    \noindent\textbf{Algorithm \thealgorithm.} #1
    \par\vspace{4pt}
    \hrule\vspace{4pt}
}

\begin{document}
% \preprint{HEP/123-qed}
\title{Quantum Architecture Search for Quantum Convolutional Neural Networks Using Double Deep Q-Networks and Bayesian Optimization}
\author{Li-An Lo}
\affiliation{Department of Engineering Science, National Cheng Kung University, Tainan 701401, Taiwan}
\author{Hsien-Yi Hsieh\,\orcidlink{0000-0001-5227-8248}}
\affiliation{Institute of Photonics Technologies, National Tsing Hua University, Hsinchu 30013, Taiwan}
\email{akjmo972@gapp.nthu.edu.tw}
% \author{Ray-Kuang Lee\,\orcidlink{0000-0002-7171-7274}}
% \affiliation{Institute of Photonics Technologies, National Tsing Hua University, Hsinchu 30013, Taiwan}
\author{Li-Yi Hsu}
\affiliation{Department of Physics, Chung Yuan Christian University, Chungli 32081, Taiwan}
\affiliation{Physics Division, National Center for Theoretical Sciences, Taipei 106319, Taiwan}
% \begin{IEEEkeywords}
% Bayesian optimization,
% deep reinforcement learning,
% quantum architecture search,
% quantum convolutional neural networks,
% quantum machine learning,
% two-qubit ansatzes.
% \end{IEEEkeywords}

\begin{abstract}
The performance of quantum convolutional neural networks (QCNNs) depends on the choice of parameterized quantum circuits used as their convolutional building blocks. In this work, we investigate quantum architecture search for designing two-qubit ansatzes in QCNNs using double deep Q-networks (DDQN) and Bayesian optimization (BO). While DDQN constructs candidate circuits sequentially using length-dependent adaptive accuracy baselines, BO explores complete gate sequences through a probabilistic surrogate model. Both approaches are evaluated on binary classification tasks derived from MNIST and Fashion-MNIST under angle and amplitude encoding. Although circuit complexity is not explicitly penalized during the search, thirteen of the twenty-four selected ansatzes contain only two CNOT gates, seven of which achieve higher classification accuracies than the universal $SU(4)$ reference ansatz. As an additional investigation, we examine cross-task transfer by reusing source-trained quantum convolutional layers with both their architectures and variational parameters frozen. These findings highlight the potential of automated architecture search for identifying task-dependent and compact quantum convolutional building blocks.
\end{abstract}

\date{\today}
\maketitle

\section{Introduction}

Quantum computing exploits nonclassical resources,
such as superposition and entanglement, to process
information in high-dimensional Hilbert spaces.
Quantum algorithms for integer factorization and
unstructured database search have demonstrated
the potential of quantum computation to provide
computational advantages over classical approaches
\cite{Shor,Grover}. With the rapid development of
quantum technologies and the success of classical
machine learning, quantum machine learning (QML)
has attracted considerable attention, particularly
in the noisy intermediate-scale quantum (NISQ)
era \cite{1,2,3,4,5}.

Parameterized quantum circuits (PQCs) provide
a flexible framework for constructing quantum
machine learning models. However, the optimization
of their variational parameters can be challenging.
In particular, barren plateaus, characterized by
exponentially vanishing gradient variances under
certain conditions, may hinder the training of
quantum neural networks \cite{6,7,8,9}.

Quantum convolutional neural networks (QCNNs)
\cite{10}, inspired by classical convolutional
neural networks (CCNNs), provide a structured
approach to quantum machine learning. Their
hierarchical architectures can mitigate certain
trainability problems associated with PQCs
\cite{11,8,9,14}. QCNNs have been investigated
for various applications, including image
classification and recognition
\cite{44,45,46,47,471}, damage and object
detection \cite{49,50}, and data analysis in
high-energy physics \cite{48}.

Similar to CCNNs, QCNNs employ convolutional
and pooling layers to extract relevant features
and progressively reduce the effective dimension
of the representation. In QCNNs, parameterized
quantum operations are applied to subsets of
qubits, followed by pooling operations that
reduce the number of active qubits. The
variational parameters are optimized using
classical computational resources within a
hybrid quantum--classical framework.

The design of a PQC involves a trade-off
between its representational capability,
trainability, and implementation cost.
Increasing the number of quantum gates
may enhance the expressibility of a
parameterized circuit \cite{qas3,qas4},
but can also increase its susceptibility
to hardware noise and optimization
difficulties \cite{qas1,qas2,qas5,8,qas7}.
While quantum error mitigation provides
strategies for addressing hardware
imperfections \cite{qem,qemm}, circuit
architecture design offers a complementary
approach to controlling circuit complexity.

Quantum architecture search (QAS) aims to
automatically identify suitable quantum
circuit structures for a given computational
task. Various approaches have been explored,
including genetic and evolutionary algorithms
\cite{ge1,ge2,ev1,ev2}, differentiable
architecture search \cite{wei}, and
training-free or graph-based search
strategies \cite{free,free1}.

In QCNN architectures constructed from
local two-qubit ansatzes, parameterized
quantum circuits serve as fundamental
building blocks of the convolutional
layers \cite{15,18,181,21}.
The choice of these ansatzes determines
the local quantum transformations used
throughout the network and can influence
its classification performance and
circuit complexity.

Previous studies have investigated
ansatz-based architecture search for
variational quantum eigensolvers
\cite{qas2,ge1,bu2} and quantum approximate
optimization algorithms \cite{bu3}.
In QCNNs, however, the hierarchical
architecture provides a natural
opportunity to perform architecture
search at the level of individual
two-qubit building blocks.

Rather than searching over the complete
multi-qubit circuit, we focus on the
design of these local ansatzes while
preserving the overall convolutional
architecture. This reduces the scope
of the architecture-search problem
and allows us to investigate how
task-dependent local circuit designs
influence classification performance
and entangling-gate complexity.

Various two-qubit circuit designs have
been employed in QCNNs for image
classification \cite{15,q11,q5},
multi-nutrient detection \cite{49},
unsupervised feature extraction \cite{21},
and transfer learning \cite{16}.
The expressibility and entangling
capabilities of parameterized two-qubit
circuits have also been extensively
investigated \cite{18,15,an1,an2,an3,119,20}.

From a circuit-design perspective,
an arbitrary two-qubit unitary operation
can be decomposed into at most three
CNOT gates together with single-qubit
unitary operations \cite{119,20}.
However, a particular learning task
may not require the full representational
capability of a universal two-qubit
ansatz. This motivates the investigation
of task-dependent circuit structures
that retain competitive classification
performance with relatively few
entangling gates.

An important question is therefore
whether an accuracy-driven architecture
search can identify suitable two-qubit
ansatzes for different classification
tasks without relying exclusively on
predefined circuit designs. In particular,
we investigate whether compact ansatzes
containing only two CNOT gates can
be identified without explicitly
introducing circuit-complexity penalties
into the search objective.

Reinforcement learning (RL) and Bayesian
optimization (BO) provide two alternative
approaches to exploring the quantum
architecture space. In RL-based QAS,
circuit construction can be formulated
as a sequential decision process, in
which an agent selects quantum gates
according to feedback obtained from
candidate circuits. Such approaches have
been investigated for quantum state
preparation \cite{rl1,rl2}, quantum
compilation \cite{rl3}, and variational
quantum circuit optimization \cite{rl4}.

In contrast, BO employs a probabilistic
surrogate model to estimate the performance
of candidate architectures and guide the
selection of subsequent circuit evaluations.
Recent studies have explored representation
learning for Bayesian quantum architecture
search \cite{bo2}, as well as its application
to variational quantum algorithms \cite{bo3}.
These developments motivate the investigation
of both sequential and surrogate-guided
search strategies for designing quantum
machine learning architectures.

In this work, we investigate DQN- and
BO-based quantum architecture search
for designing two-qubit ansatzes in
QCNNs. Specifically, the double deep
Q-network (DDQN) variant is employed
to construct candidate circuits
sequentially from elementary quantum gates.
For brevity, we use DQN to denote the
DDQN-based search method in the
experimental comparisons.

The DDQN-based search incorporates
length-dependent adaptive accuracy
baselines and an experience-replay
mechanism in which rewards are
recomputed from stored classification
accuracies. In contrast, BO explores
complete candidate gate sequences
using a Gaussian-process surrogate
model with a Mat\'ern covariance kernel.

The main contributions of this work
are summarized as follows:

\begin{enumerate}
    \item We investigate two complementary
    architecture-search strategies for
    designing task-dependent two-qubit
    ansatzes while preserving the overall QCNN architecture. The DDQN-based
    approach incorporates length-dependent
    adaptive accuracy baselines and
    reward recomputation during
    experience replay.

    \item We systematically evaluate the
    searched ansatzes across multiple
    binary classification tasks under
    angle and amplitude encoding, and
    compare their classification
    performance with that of a universal
    two-qubit reference ansatz.

    \item We identify compact two-qubit
    ansatzes containing only two CNOT
    gates, some of which achieve
    competitive or improved classification
    accuracies without explicitly
    penalizing circuit complexity
    during architecture search.

    \item We further investigate cross-task
    and cross-dataset reusability by
    transferring both the architecture
    and optimized variational parameters
    of a source-trained first convolutional
    layer and keeping them frozen during
    target-task training.
\end{enumerate}

The remainder of this paper is organized
as follows. Section II reviews the
QCNN architecture and the relevant
background of the architecture-search
methods. Section III introduces the
data-embedding strategies and the
classification objective. Section IV
describes the DQN- and BO-based
architecture-search procedures and
the cross-task transfer protocol.
Section V presents the experimental
results, and Sec.~VI discusses the
findings, limitations, and future
research directions.

\section{Related Work}

\subsection{QCNN architecture and two-qubit ansatzes}

In this work, we consider a QCNN
architecture consisting of alternating
quantum convolutional and pooling layers.
Within each convolutional layer, identical
parameterized two-qubit circuit architectures
are applied to neighboring qubit pairs
in a translationally invariant manner
with periodic boundary conditions.
The pooling layers progressively reduce
the number of active qubits.

Several parameterized two-qubit circuit
designs have been investigated in
Refs.~\cite{15,16}. These architectures
differ in their expressibility,
entangling capability, and circuit
complexity. In particular, parameterized
$SU(4)$ circuits provide a universal
two-qubit ansatz and have been widely
employed in QCNN architectures
\cite{15,16,18,181,21}.

An arbitrary two-qubit unitary, up to
a global phase, can be decomposed into
at most three CNOT gates and single-qubit
unitary operations \cite{20}.
As our reference ansatz, we employ
the universal two-qubit circuit used
in our previous work \cite{Lo}.
This ansatz consists of three CNOT gates
and single-qubit unitary operations with
a total of 15 trainable parameters.
It is denoted as Original in the
experimental comparisons.

For the considered $n$-qubit QCNN
architecture with periodic nearest-neighbor
connectivity, the first convolutional layer
contains $n$ two-qubit convolutional blocks.
Using the reference ansatz therefore
requires $3n$ CNOT gates in this layer.
Replacing it with a two-CNOT ansatz
reduces this count to $2n$, corresponding
to a reduction of one third.

These considerations motivate the search
for task-dependent two-qubit ansatzes
using DQN and Bayesian optimization,
as introduced in the following subsections.

\subsection{Deep Q-networks}

Reinforcement learning (RL) provides
a framework in which an agent interacts
with an environment and learns a policy
through the rewards received from its
actions. At each step, the agent
observes a state $s$, selects an action
$a$, receives a reward $r$, and
transitions to a new state $s'$.

In $Q$-learning, the action-value
function $Q^\pi(s,a)$ quantifies the
expected discounted cumulative reward
obtained by taking action $a$ in state
$s$ and subsequently following policy
$\pi$. The optimal action-value
function satisfies the Bellman
optimality equation
\begin{equation}
Q^{*}(s,a)
=
\mathbb{E}
\left[
r+\gamma\max_{a'}Q^{*}(s',a')
\mid s,a
\right],
\end{equation}
where $\gamma$ denotes the discount
factor.

For problems with large state spaces,
explicitly storing the action-value
function in a $Q$-table becomes
impractical. Deep Q-networks (DQNs)
address this limitation by approximating
the action-value function using a
neural network,
\begin{equation}
Q(s,a)\simeq Q(s,a;\boldsymbol{w}),
\end{equation}
where $\boldsymbol{w}$ denotes the
trainable network parameters.

The network parameters are optimized
by minimizing the temporal-difference
error between the predicted action
value and a target value. In the
standard DQN formulation, the target
for a nonterminal transition is
given by
\begin{equation}
y
=
r+\gamma\max_{a'}
Q(s',a';\boldsymbol{w}^{-}),
\end{equation}
where $\boldsymbol{w}^{-}$ denotes
the parameters of a target network.
For terminal transitions, the target
reduces to the immediate reward $r$.

Experience replay and a separate
target network are commonly employed
to improve the stability of DQN
training. DQNs and their variants
have been applied to various
sequential decision-making problems,
including game playing and control
tasks \cite{221,222,223,225}.

The sequential nature of RL provides
a natural formulation for quantum
architecture search. A partially
constructed quantum circuit can
be regarded as the current state,
while the selection of the next
quantum gate corresponds to an
action. The performance of the
resulting parameterized quantum
circuit provides feedback for
constructing the reward.

In the present work, this formulation
is employed to construct candidate
two-qubit ansatzes for QCNNs.
The specific state representation,
action space, reward formulation,
and DQN implementation are described
in Sec.~IV.

\subsection{Bayesian optimization}

Bayesian optimization (BO) is a global
optimization strategy designed for
objective functions that are expensive
to evaluate and for which gradient
information may be unavailable.
Instead of exhaustively evaluating
the objective function over the
entire search space, BO constructs
a probabilistic surrogate model
from previously evaluated candidates
and uses this model to guide subsequent
evaluations \cite{b1,b3,bo3}.

This framework is particularly relevant
to quantum architecture search,
where evaluating a candidate architecture
requires training and evaluating the
corresponding parameterized quantum
circuit.

A Gaussian process (GP) is commonly
employed as the surrogate model.
For an unknown objective function
$f(\boldsymbol{x})$, a GP prior
is written as
\begin{equation}
f(\boldsymbol{x})
\sim
\mathcal{GP}
\left[
m(\boldsymbol{x}),
k(\boldsymbol{x},\boldsymbol{x}')
\right],
\end{equation}
where $m(\boldsymbol{x})$ denotes
the mean function and
$k(\boldsymbol{x},\boldsymbol{x}')$
denotes the covariance kernel.

Given previously evaluated candidates
and their corresponding objective
values, the GP posterior provides
a predictive mean and uncertainty
for unevaluated candidates.

An acquisition function is used
to guide the selection of the next
candidate by balancing exploitation
of candidates with high predicted
performance and exploration of
regions with large predictive
uncertainty.

One commonly used acquisition
function is expected improvement
(EI). For a maximization problem,
it is defined as
\begin{equation}
\mathrm{EI}(\boldsymbol{x})
=
\mathbb{E}
\left[
\max\left(
f(\boldsymbol{x})-f_{\mathrm{best}},0
\right)
\right],
\end{equation}
where $f_{\mathrm{best}}$ denotes
the best objective value observed
so far.

Under a Gaussian posterior with
predictive mean $\mu(\boldsymbol{x})$
and standard deviation
$\sigma(\boldsymbol{x})>0$,
the expected improvement can be
expressed as
\begin{equation}
\mathrm{EI}(\boldsymbol{x})
=
\left[
\mu(\boldsymbol{x})-f_{\mathrm{best}}
\right]\Phi(Z)
+
\sigma(\boldsymbol{x})\phi(Z),
\end{equation}
where
\begin{equation}
Z=
\frac{
\mu(\boldsymbol{x})-f_{\mathrm{best}}
}{
\sigma(\boldsymbol{x})
},
\end{equation}
and $\Phi(\cdot)$ and $\phi(\cdot)$
denote the cumulative distribution
function and probability density
function of the standard normal
distribution, respectively.

In contrast to the sequential
gate-selection mechanism of DQN,
BO evaluates complete candidate
architectures. Previously evaluated gate sequences and their corresponding classification
accuracies are used to construct a surrogate model that guides the selection of subsequent candidates.

The representation of candidate
architectures and the GP surrogate
model employed in the present
work are described in Sec.~IV.

\section{Data embedding and loss function}
\label{sec:data_embedding}

To apply QCNNs to classical
image-classification tasks, classical
input features must first be mapped
onto quantum states. In this work,
we consider two commonly used
quantum data-embedding strategies:
amplitude encoding and angle
encoding \cite{23}.

Let
\begin{equation}
\boldsymbol{x}
=
(x_1,x_2,\ldots,x_N)
\in\mathbb{R}^{N}
\end{equation}
denote an $N$-dimensional classical
feature vector to be encoded into
an $n$-qubit quantum system.

The two encoding strategies differ
in their feature-to-qubit mappings,
state-preparation procedures, and
qubit requirements. Consequently,
the choice of data embedding may
influence the suitability of a
particular two-qubit ansatz for
the subsequent classification task.

\subsection{Amplitude encoding}

Amplitude encoding maps the
components of a nonzero classical
feature vector $\boldsymbol{x}$
onto the probability amplitudes
of a quantum state,
\begin{equation}
|\psi(\boldsymbol{x})\rangle
=
\frac{1}{\|\boldsymbol{x}\|_2}
\sum_{i=1}^{N}x_i|i-1\rangle,
\label{eq:amplitude_encoding}
\end{equation}
where
\begin{equation}
\|\boldsymbol{x}\|_2
=
\sqrt{\sum_{i=1}^{N}|x_i|^2},
\end{equation}
and $|i-1\rangle$ denotes the
corresponding computational basis
state.

The normalization ensures that
\begin{equation}
\langle\psi(\boldsymbol{x})
|\psi(\boldsymbol{x})\rangle=1.
\end{equation}

An $N$-dimensional feature vector
can be represented using
\begin{equation}
n=\left\lceil\log_2 N\right\rceil
\end{equation}
qubits, assuming $N\geq 2$.
When $N$ is not a power of two,
the feature vector can be padded
with zeros to match the dimension
of the quantum Hilbert space.

Although amplitude encoding provides
a logarithmic qubit requirement
with respect to the feature
dimension, preparing an arbitrary
amplitude-encoded state generally
requires a nontrivial state-preparation
circuit. This may increase circuit
depth and implementation costs
on near-term quantum hardware
\cite{24}.

\subsection{Angle encoding}

Angle encoding maps individual
components of a classical feature
vector onto the rotation angles
of single-qubit quantum gates.
For a rotation about the $y$ axis,
the encoding operation is given by
\begin{equation}
R_y(x_i)
=
\exp\left(-i\frac{x_i}{2}\sigma_y\right),
\end{equation}
where $\sigma_y$ denotes the
Pauli-$Y$ operator.

Acting on the initial state
$|0\rangle$, the rotation prepares
the single-qubit state
\begin{equation}
|\phi(x_i)\rangle
=
R_y(x_i)|0\rangle
=
\cos\left(\frac{x_i}{2}\right)|0\rangle
+
\sin\left(\frac{x_i}{2}\right)|1\rangle.
\label{eq:angle_single}
\end{equation}

The complete feature vector is
encoded into an $N$-qubit
product state,
\begin{equation}
|\psi(\boldsymbol{x})\rangle
=
\bigotimes_{i=1}^{N}
|\phi(x_i)\rangle.
\label{eq:angle_encoding}
\end{equation}

Unlike amplitude encoding, this
form of angle encoding requires
one qubit for each input feature,
such that $n=N$. However, the
encoding circuit consists only
of single-qubit rotations and
does not require entangling
gates for state preparation.

Amplitude and angle encoding
therefore provide different
trade-offs between the number
of required qubits and the
complexity of state preparation.
They also generate different
quantum representations of
classical data, which may
influence the performance of
the searched two-qubit ansatzes.

\subsection{Loss function}

Let $A$ denote a candidate two-qubit
ansatz architecture and let
$\boldsymbol{\theta}$ denote the
trainable variational parameters
of the corresponding QCNN.

For binary classification, the
measurement results obtained from
the QCNN are used to determine
the predicted probability
\begin{equation}
p(\boldsymbol{x};\boldsymbol{\theta},A)
\in [0,1]
\end{equation}
associated with class $1$.

Given a training dataset
\begin{equation}
\mathcal{D}_{\mathrm{tr}}
=
\left\{
(\boldsymbol{x}_j,y_j)
\right\}_{j=1}^{N_{\mathrm{tr}}},
\qquad
y_j\in\{0,1\},
\end{equation}
where $N_{\mathrm{tr}}$ denotes the
number of training samples,
the variational parameters are
optimized by minimizing the binary
cross-entropy loss
\begin{equation}
\mathcal{L}(\boldsymbol{\theta};A)
=
-\frac{1}{N_{\mathrm{tr}}}
\sum_{j=1}^{N_{\mathrm{tr}}}
\left[
y_j\log p_j+(1-y_j)\log(1-p_j)
\right],
\label{eq:bce_loss}
\end{equation}
where
\begin{equation}
p_j
=
p(\boldsymbol{x}_j;
\boldsymbol{\theta},A).
\end{equation}

For a fixed architecture $A$, the variational
parameters are numerically optimized by
minimizing the BCE loss on the training set.
The resulting parameters are denoted by
$\boldsymbol{\theta}^{*}(A)$.

The classification accuracy associated with
architecture $A$ is defined as
\begin{equation}
\mathcal{A}(A)
=
\mathrm{Acc}_{\mathrm{val}}
\left(A,\boldsymbol{\theta}^{*}(A)\right),
\label{eq:architecture_accuracy}
\end{equation}
where $\mathrm{Acc}_{\mathrm{val}}$ denotes
the classification accuracy evaluated on
the validation set.

This validation accuracy serves as the
performance criterion for both DDQN- and
BO-based architecture search.

It is important to distinguish
the optimization of variational
parameters from the optimization
of the circuit architecture.
The former minimizes the
classification loss for a fixed
gate sequence, whereas the latter
searches over different gate
sequences according to their
classification performance.

The architecture-search problem
can therefore be expressed
schematically as
\begin{equation}
A^{*}
\approx
\arg\max_{A\in\mathcal{S}}
\mathcal{A}(A),
\label{eq:architecture_objective}
\end{equation}
where $\mathcal{S}$ denotes the
space of candidate two-qubit
ansatz architectures.

In the present work, DQN and
Bayesian optimization are used
to explore this discrete
architecture space, as described
in Sec.~IV.

\section{Methods}
\label{sec:methods}

In this section, we describe the quantum
architecture search (QAS) strategies used
to construct the two-qubit ansatzes of the QCNN.
Rather than selecting a two-qubit circuit
from a predefined ansatz library, we construct
candidate ansatzes from a set of elementary
quantum gates. Two complementary search
strategies are considered: a DQN-based
approach, which constructs the ansatz
sequentially, and Bayesian optimization (BO),
which uses a probabilistic surrogate model
to guide the selection of complete candidate
gate sequences.

The architecture search is guided by
classification performance without explicitly
penalizing the number of CNOT gates or
the total gate count. Circuit complexity
is therefore examined as an empirical
property of the searched ansatzes, with
particular attention to two-CNOT circuits
and their classification performance relative
to the reference ansatz introduced in Sec.~II.

\subsection{Quantum architecture search space}
\label{subsec:search_space}

A candidate two-qubit ansatz is represented as an ordered sequence of
elementary quantum gates,
\begin{equation}
A=(g_1,g_2,\ldots,g_L),
\label{eq:ansatz_sequence}
\end{equation}
where $g_l\in\mathcal{G}$ and $\mathcal{G}$ denotes the elementary
gate set used in the architecture search. The order of the gates is
important because, in general, quantum operations do not commute.
Consequently, different sequences constructed from the same set of
gates can represent different two-qubit transformations.

In the present implementation, the elementary
gate set consists of ten possible gate operations,
\begin{equation}
\begin{aligned}
\mathcal{G}=\big\{&
R_x^{(1)},R_x^{(2)},
R_y^{(1)},R_y^{(2)},\\
&
R_z^{(1)},R_z^{(2)},
U^{(1)},U^{(2)},\\
&
\mathrm{CNOT}_{1\rightarrow2},
\mathrm{CNOT}_{2\rightarrow1}
\big\},
\end{aligned}
\label{eq:elementary_gate_set}
\end{equation}
where the superscripts indicate the qubit
on which the single-qubit gate acts,
and $U$ denotes a general three-parameter
single-qubit unitary gate.
For numerical representation, we assign
indices $0$ through $9$ to these gate
operations, respectively, following the
listing in Eq.~\eqref{eq:elementary_gate_set}.

The corresponding architecture search space
is defined as
\begin{equation}
\mathcal{S}
=
\bigcup_{L=1}^{L_{\max}}\mathcal{G}^{L},
\label{eq:architecture_search_space}
\end{equation}
where $\mathcal{G}^{L}$ denotes the set
of ordered gate sequences of length $L$,
and $L_{\max}$ is the maximum allowed
gate sequence length.

Controlled rotation gates are not introduced
as independent search operations because
they can be decomposed into elementary
one- and two-qubit gates.
Both the total gate count and the number
of CNOT gates can therefore be determined
directly from each candidate gate sequence.

For each candidate architecture $A$,
the corresponding QCNN is trained,
and its validation accuracy $\mathcal{A}(A)$,
defined in Eq.~\eqref{eq:architecture_accuracy},
is used to guide the architecture search.

\subsection{DDQN-based architecture search}
\label{subsec:dqn_qas}

\begin{figure}[t]
    \centering
    \includegraphics[width=\columnwidth]
    {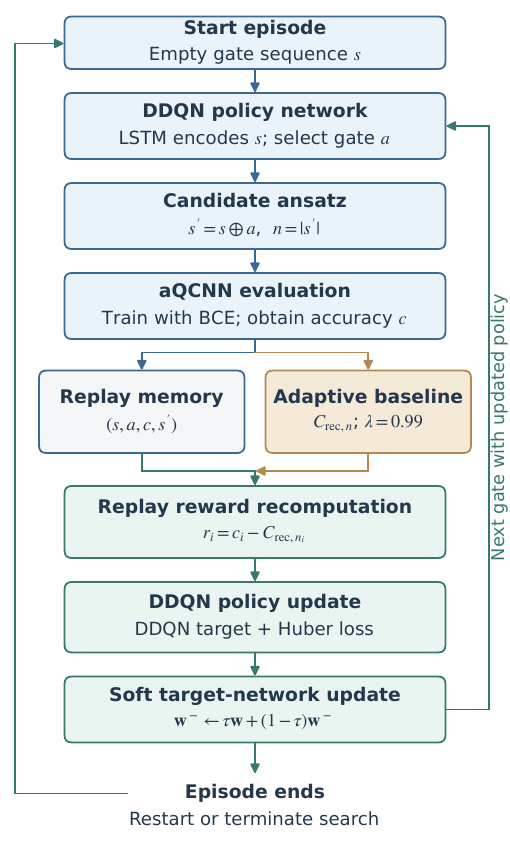}
    \caption{
    Overview of the DDQN-based quantum
    architecture search with adaptive
    accuracy baselines.
    The DDQN agent is updated after
    each gate addition using sampled
    experiences and recomputed rewards.
    }
    \label{fig:ddqn_framework}
\end{figure}

An overview of the DDQN-based
architecture-search framework is
illustrated in Fig.~\ref{fig:ddqn_framework}.
The procedure consists of sequential
ansatz construction, classification-based
evaluation, and adaptive experience
replay for updating the DDQN agent.

The architecture search is organized into
multiple episodes, each starting from an
empty gate sequence. Within each episode,
the ansatz is constructed sequentially by
appending gates selected by the policy
network. Each search episode is subject to
a prescribed maximum gate sequence length.

The construction of a two-qubit ansatz can
naturally be formulated as a sequential
decision process. At each gate-selection step,
the state of the environment is defined by
the partially constructed gate sequence,
\begin{equation}
s=(g_1,g_2,\ldots,g_{\ell}),
\label{eq:dqn_state}
\end{equation}
where $\ell$ denotes the current gate
sequence length.
The action $a$ corresponds to selecting
the next gate from the available gate set,
\begin{equation}
a\in\mathcal{G},
\end{equation}
such that the next state becomes
\begin{equation}
s'=s\oplus a
=(g_1,g_2,\ldots,g_{\ell},a).
\label{eq:dqn_transition}
\end{equation}
Here, $\oplus$ denotes the operation of
appending a gate to the current sequence.
For an initially empty ansatz, a special
token is used as the initial input to
the sequence model.

Because the state is an ordered gate
sequence with a variable length,
a long short-term memory (LSTM) network
is employed to encode the current
architecture. The resulting representation
is used by the DQN to estimate the
action values
\begin{equation}
Q(s,a;\boldsymbol{w}),
\end{equation}
where $\boldsymbol{w}$ denotes the
trainable parameters of the DQN.
The quantum gate selected by the agent
is subsequently appended to the current
ansatz, and the resulting circuit is
evaluated based on the classification
accuracy of the corresponding QCNN.

The reward is defined using a length-dependent
adaptive accuracy baseline.
Let $n_a$ denote the number of gate operations
in the candidate ansatz, $t$ denote the
search episode index, and $k$ denote the
global search step, which increases after
each gate addition.
For each sequence length $n_a$, we maintain
a recorded classification accuracy
$C_{\mathrm{rec},n_a}^{t}$.

The recorded accuracy is updated according to
\begin{equation}
C_{\mathrm{rec},n_a}^{t+1}
=
\begin{cases}
C_{\mathrm{now},n_a}^{t},
& C_{\mathrm{now},n_a}^{t}
\geq C_{\mathrm{rec},n_a}^{t},\\[4pt]
0.99C_{\mathrm{rec},n_a}^{t},
& \text{otherwise},
\end{cases}
\label{eq:recorded_accuracy_update}
\end{equation}
where $C_{\mathrm{now},n_a}^{t}$ denotes
the classification accuracy of the candidate
ansatz of length $n_a$ evaluated in episode $t$. The superscript $t$ indexes the
episode-wise updates of each baseline,
whereas the superscript $(k)$ denotes
the current value of the same baseline
at global search step $k$. The decay mechanism is intended to mitigate the influence of unusually high accuracies encountered during early search episodes, which could otherwise lead to persistently negative rewards for subsequent candidates of the same sequence length.

Unlike conventional experience replay,
which stores the immediate reward,
the replay memory in the present
architecture-search procedure stores
the classification accuracy of each
evaluated candidate ansatz.
Each transition is represented as
\begin{equation}
(s_i,a_i,c_i,s_i'),
\label{eq:replay_transition}
\end{equation}
where $c_i$ denotes the classification
accuracy associated with the candidate
architecture $s_i'$.

After each gate addition, the candidate
QCNN is trained and evaluated, and the
resulting transition is stored in the
replay memory.
The recorded accuracy vector is updated
according to Eq.~\eqref{eq:recorded_accuracy_update}
before a mini-batch of stored transitions
is sampled to update the policy network.
During experience replay, the reward
is computed relative to the updated
length-dependent accuracy baseline,
\begin{equation}
r_i^{(k)}
=
c_i-C_{\mathrm{rec},n_i}^{(k)},
\label{eq:replay_reward}
\end{equation}
where $n_i=|s_i'|$ denotes the gate sequence
length of the sampled candidate architecture,
and $C_{\mathrm{rec},n_i}^{(k)}$ denotes
the corresponding accuracy baseline retrieved
from the updated recorded accuracy vector
at global search step $k$.
Consequently, the reward associated
with a stored transition may change
as the recorded accuracy is updated
during architecture search.

Following the double DQN formulation,
action selection and action-value
evaluation are performed using the
policy and target networks, respectively.
For a nonterminal transition, the
temporal-difference target is given by
\begin{equation}
\begin{aligned}
y_i^{(k)}
={}&r_i^{(k)}\\
&+\gamma Q\left(
s_i',
\arg\max_{a'}Q(s_i',a';\boldsymbol{w});
\boldsymbol{w}^{-}
\right),
\end{aligned}
\label{eq:ddqn_target}
\end{equation}
where $\gamma$ is the discount factor,
and $\boldsymbol{w}$ and
$\boldsymbol{w}^{-}$ denote the
parameters of the policy and target
networks, respectively.

The temporal-difference error is defined as
\begin{equation}
\delta_i^{(k)}
=
y_i^{(k)}
-
Q(s_i,a_i;\boldsymbol{w}).
\label{eq:td_error}
\end{equation}

The policy-network parameters are
optimized by minimizing the Huber loss
\begin{equation}
\mathcal{L}_{\mathrm{DDQN}}
=
\mathbb{E}_{i\sim\mathcal{B}}
\left[
h_{\kappa}(\delta_i^{(k)})
\right],
\label{eq:dqn_loss}
\end{equation}
where $\mathcal{B}$ denotes a mini-batch
sampled from the replay memory, and
\begin{equation}
h_{\kappa}(\delta)
=
\begin{cases}
\frac{1}{2}\delta^2,
& |\delta|\leq\kappa,\\
\kappa(|\delta|-\frac{1}{2}\kappa),
& \text{otherwise},
\end{cases}
\label{eq:huber_loss}
\end{equation}
with $\kappa$ denoting the Huber threshold.
The Huber loss is quadratic for small
temporal-difference errors and linear
for large errors.

The target network is updated using
a soft-update rule,
\begin{equation}
\boldsymbol{w}^{-}
\leftarrow
\tau\boldsymbol{w}
+(1-\tau)\boldsymbol{w}^{-},
\label{eq:soft_target_update}
\end{equation}
where $\tau$ is the soft-update
coefficient.

Importantly, candidate ansatzes are not
restricted to the circuits obtained at the
last step of each episode. Instead, the
intermediate gate sequences generated during
the search are also evaluated. Let
\begin{equation}
A_k=(g_1,g_2,\ldots,g_{L_k})
\end{equation}
denote the candidate architecture evaluated
at global search step $k$, where $L_k$ is
its gate sequence length.

The evaluated architectures and their
corresponding validation accuracies are
recorded throughout the search. Upon
completion of the search, high-performing
candidate architectures are retained for
further evaluation, as described in
Sec.~\ref{subsec:ansatz_selection}.

Consequently, the final selected ansatz
may contain fewer gate operations than
the maximum allowed sequence length.
Such variable-length solutions arise
from accuracy-driven architecture search
rather than explicit minimization of
the number of quantum gates or CNOT
operations.

\subsection{Bayesian-optimization-based architecture search}
\label{subsec:bo_qas}

As an alternative to the sequential construction used by the DQN, we
also employ Bayesian optimization to search over complete candidate
architectures. To obtain a fixed-dimensional representation required
by the surrogate model, each candidate ansatz is represented by a
sequence of fixed maximum length,
\begin{equation}
\widetilde{A}
=
(\widetilde{g}_1,\widetilde{g}_2,\ldots,\widetilde{g}_{L_{\max}}),
\label{eq:bo_representation}
\end{equation}
where each nonempty element specifies a gate operation and a special
null value is used to indicate an unused position. For example, a
representation of the form
\begin{equation}
(1,-1,8,0,-1,2,-1,-1,-1,-1)
\end{equation}
corresponds to the effective gate sequence
\begin{equation}
(1,8,0,2).
\end{equation}

Let $A_i$ denote the effective gate sequence
obtained by removing the null entries from
$\widetilde{A}_i$.
After evaluating an initial set of candidate
architectures, the observations
\begin{equation}
\mathcal{D}_{\mathrm{BO}}
=
\left\{
\left(
\widetilde{A}_i,
\mathcal{A}(A_i)
\right)
\right\}_{i=1}^{N_{\mathrm{obs}}}
\label{eq:bo_dataset}
\end{equation}
are used to construct a Gaussian-process
surrogate model. A Mat\'ern covariance kernel
is employed to model the relation between
the encoded architecture and its
classification performance.

Based on the posterior predictive distribution of the surrogate model,
an acquisition function is evaluated to determine the next candidate
architecture. The selected architecture is used to construct and train
the corresponding QCNN, after which its classification accuracy is
added to $\mathcal{D}_{\mathrm{BO}}$. The Gaussian-process posterior
is then updated, and the procedure is repeated. In contrast to the
DQN, which learns a policy for selecting the next gate conditioned on
a partially constructed circuit, BO directly proposes complete
candidate architectures based on the accumulated architecture-
performance observations.

\subsection{Architecture selection and performance evaluation}
\label{subsec:ansatz_selection}

Both architecture-search procedures evaluate
candidate two-qubit ansatzes according to
their validation accuracies. To reduce the
influence of stochastic parameter optimization
on architecture selection, a two-stage
selection procedure is employed.

Upon completion of the architecture search,
the top $M$ candidate ansatzes are retained
according to their validation accuracies.
Let $\mathcal{S}_M$ denote the resulting
set of shortlisted architectures.

For each candidate $A\in\mathcal{S}_M$,
the corresponding QCNN is retrained $R$
times using different random initializations
of its variational parameters.
The mean validation accuracy is defined as
\begin{equation}
\overline{\mathcal{A}}_{\mathrm{val}}(A)
=
\frac{1}{R}
\sum_{r=1}^{R}
\mathrm{Acc}_{\mathrm{val}}
\left(A,\boldsymbol{\theta}_{A}^{(r)}\right),
\label{eq:mean_validation_accuracy}
\end{equation}
where $\boldsymbol{\theta}_{A}^{(r)}$
denotes the optimized variational parameters
obtained from the $r$th training run.

The final ansatz architecture is selected
according to
\begin{equation}
A^*
=
\underset{A\in\mathcal{S}_M}{\arg\max}
\overline{\mathcal{A}}_{\mathrm{val}}(A).
\label{eq:final_architecture_selection}
\end{equation}

After architecture selection, the $R$
trained QCNN models corresponding to
$A^*$ are evaluated on the held-out
test set. The reported classification
accuracy is defined as
\begin{equation}
\overline{\mathcal{A}}_{\mathrm{test}}(A^*)
=
\frac{1}{R}
\sum_{r=1}^{R}
\mathrm{Acc}_{\mathrm{test}}
\left(A^*,\boldsymbol{\theta}_{A^*}^{(r)}\right).
\label{eq:mean_test_accuracy}
\end{equation}

Importantly, the test set is not used
during architecture search or architecture
selection.

Circuit complexity is evaluated separately
in terms of the total gate count and
the number of CNOT gates in each
two-qubit ansatz.

\subsection{Cross-task transfer of ansatzes}
\label{subsec:architecture_transfer}

In addition to task-specific architecture search, we
investigate the reuse of ansatzes across different
classification tasks. In the standard setting, the same
searched two-qubit ansatz architecture
is used across the convolutional layers
of the QCNN. In the
cross-task experiments~\cite{Lo}, this constraint is relaxed
by introducing a source-task ansatz into the first
convolutional layer.

Let $A_{\mathrm{S}}$ denote the ansatz architecture
obtained from a source classification task, and
let $\boldsymbol{\theta}_{\mathrm{S}}$ denote its
optimized variational parameters. The first
convolutional layer of the target model is
constructed using the transferred pair
\begin{equation}
\left(
A_{\mathrm{T}}^{(1)},
\boldsymbol{\theta}_{\mathrm{T}}^{(1)}
\right)
=
\left(
A_{\mathrm{S}},
\boldsymbol{\theta}_{\mathrm{S}}
\right),
\label{eq:ansatz_parameter_transfer}
\end{equation}
where the superscript denotes the convolutional
layer index. Both the transferred ansatz architecture
and its variational parameters are kept fixed
throughout target-task training. In particular,
the transferred parameters
$\boldsymbol{\theta}_{\mathrm{S}}$ are frozen
and are not updated using the target-task data.

In contrast, the remaining convolutional
layers are constructed for the target
classification task. Their ansatz
architectures are either obtained using
the selected architecture-search strategy
or specified by the reference ansatz,
depending on the experimental configuration.
Their variational parameters are optimized
using the target-task data.
Thus, only the parameters of the
non-transferred layers are updated during
target-task training. Schematically, the resulting target model is represented by
\begin{equation}
\mathcal{M}_{\mathrm{T}}
=
\left[
(A_{\mathrm{S}},\boldsymbol{\theta}_{\mathrm{S}}),
(A_{\mathrm{T}}^{(2)},\boldsymbol{\theta}_{\mathrm{T}}^{(2)}),
(A_{\mathrm{T}}^{(3)},\boldsymbol{\theta}_{\mathrm{T}}^{(3)})
\right].
\label{eq:cross_task_model}
\end{equation}

We consider both within-dataset and cross-dataset
transfer settings, including transfer between
different MNIST classification tasks and transfer
from Fashion-MNIST to MNIST. The purpose of these
experiments is to examine the classification
performance obtained when a source-task ansatz
and its learned parameters are incorporated into
a target-task model.

The transfer experiments are treated as an
additional evaluation of the architecture-search
framework rather than as evidence of improved
search efficiency or universally beneficial
transfer learning.
\bigskip

\section{Experimental Results}
\label{sec:results}

In this section, we evaluate the performance of
two-qubit ansatzes identified using DQN and Bayesian optimization (BO) for binary
image-classification tasks. The searched ansatzes
are compared with the universal two-qubit reference
ansatz introduced in Sec.~II.

We first investigate classification performance
across different tasks and data-embedding strategies.
We then examine the circuit structures obtained
from the accuracy-driven search, with particular
attention to ansatzes containing only two CNOT
gates. Finally, we consider cross-task transfer
experiments in which a source-trained first
convolutional layer is transferred and frozen
while the remaining layers are trained for
the target classification task.

\subsection{Experimental settings}
\label{subsec:experimental_settings}

The experiments are conducted on binary
classification tasks constructed from the
MNIST \cite{MNIST} and Fashion-MNIST
\cite{FashionMNIST} datasets. For MNIST, we consider
the digit pairs $(0,1)$, $(5,7)$, $(1,2)$,
and $(0,8)$. For Fashion-MNIST, we consider
the class pairs $(0,1)$ and $(1,2)$.

For angle encoding, the original
$28\times28$ images are compressed into
eight-dimensional feature vectors using
an autoencoder (AE), following the
approach described in Ref.~\cite{21}.

For amplitude encoding, the original
images are resized to $16\times16$ pixels
and subsequently flattened into
256-dimensional vectors.

The resulting classical data are encoded
into eight-qubit quantum states using
the corresponding encoding strategies
described in Sec.~III.

The QCNN consists of eight qubits, with the
searched two-qubit ansatz serving as the building
block of its convolutional layers. In the standard
architecture-search experiments, the same
two-qubit ansatz structure is used across
the convolutional layers.

For the numerical optimization of the classifier,
a batch size of $16$ and a learning rate of
$0.004$ are used.

Both DDQN and BO were executed under
a wall-clock time limit of 24 hours. For DDQN, the maximum number of search episodes was set to 3200, with a maximum gate sequence length of 10. In practice, approximately 60 episodes were completed within the time limit, corresponding to approximately 600 candidate architecture evaluations.

Under the same time constraint, BO
evaluated approximately 400 candidate
architectures, which typically contained
longer gate sequences than those evaluated
by DDQN.

For each classification task and encoding
strategy, DQN and BO are employed to identify
candidate two-qubit ansatzes. Their classification accuracies are compared with those obtained using the reference ansatz. For each classification task, a validation
set of 400 samples is extracted from the
original training dataset using a fixed
random seed. The remaining training data
are used to optimize the variational
parameters of the QCNN.

The validation set is used to evaluate
candidate architectures throughout both
DDQN- and BO-based architecture search.

Following the selection procedure
described in Sec.~\ref{subsec:ansatz_selection},
we retain the top $M=20$ candidate
ansatzes and retrain each candidate
$R=5$ times using different random
initializations.

The architecture with the highest mean
validation accuracy is selected, and
the corresponding five trained models
are evaluated on the held-out test set
to obtain the final mean test accuracy.

\subsection{Architecture search on individual classification tasks}
\label{subsec:individual_search}

We first evaluate the classification performance
of the two-qubit ansatzes identified by DQN
and BO on individual binary classification tasks.
Table~\ref{tab:individual_results} compares
their mean test accuracies with those of the
reference ansatz across six binary classification
tasks under angle and amplitude encoding.

\begin{table}[htbp]
\caption{
Mean test classification accuracies (\%)
over five training runs for the ansatzes
identified by DQN and Bayesian optimization
(BO), compared with the reference ansatz.
Original denotes the universal $SU(4)$
reference ansatz with three CNOT gates.
A superscript $\star$ denotes a searched
ansatz containing only two CNOT gates.
}
\label{tab:individual_results}
\begin{ruledtabular}
\begin{tabular}{lccc}
Dataset and encoding & DQN & BO & Original \\
\hline
MNIST $(0,1)$, Angle
    & $97.65^{\star}$ & $94.37$ & $98.24$ \\
MNIST $(0,1)$, Amplitude
    & $99.25$ & $93.00^{\star}$ & $99.32$ \\
MNIST $(5,7)$, Angle
    & $97.63$ & $95.00$ & $96.37$ \\
MNIST $(5,7)$, Amplitude
    & $89.12^{\star}$ & $87.06$ & $86.44$ \\
MNIST $(1,2)$, Angle
    & $98.31$ & $97.38$ & $97.63$ \\
MNIST $(1,2)$, Amplitude
    & $95.75$ & $97.00$ & $94.38$ \\
MNIST $(0,8)$, Angle
    & $90.79$ & $91.88$ & $91.94$ \\
MNIST $(0,8)$, Amplitude
    & $94.56^{\star}$ & $93.19^{\star}$ & $94.50$ \\
Fashion-MNIST $(0,1)$, Angle
    & $91.93^{\star}$ & $92.12^{\star}$ & $90.63$ \\
Fashion-MNIST $(0,1)$, Amplitude
    & $95.38^{\star}$ & $94.25^{\star}$ & $97.75$ \\
Fashion-MNIST $(1,2)$, Angle
    & $97.13^{\star}$ & $96.31^{\star}$ & $96.25$ \\
Fashion-MNIST $(1,2)$, Amplitude
    & $97.38^{\star}$ & $98.44^{\star}$ & $97.44$
\end{tabular}
\end{ruledtabular}
\end{table}

As shown in Table~\ref{tab:individual_results},
the searched architectures achieve higher
classification accuracies than the reference
ansatz in several settings, although the
improvements are not uniform across tasks
and encoding strategies.

For MNIST $(5,7)$ with angle encoding, the ansatz identified by DQN achieves an accuracy of
$97.63\%$, compared with $96.37\%$ for the
reference ansatz, corresponding to an improvement
of $1.26$ percentage points. For MNIST $(1,2)$
with amplitude encoding, BO achieves $97.00\%$,
exceeding the reference accuracy of $94.38\%$
by $2.62$ percentage points.

A similar dependence is observed for
Fashion-MNIST. For the $(0,1)$ task,
both search methods outperform the
reference ansatz under angle encoding
but perform worse under amplitude encoding.
These observations suggest that the
effectiveness of a particular two-qubit
ansatz depends on both the classification
task and the data-embedding strategy.

The results also demonstrate that the relative
performance of DQN and BO depends on the
classification task. Across the twelve task--encoding combinations, DQN and BO outperform
the reference ansatz in seven and five cases, respectively. However, neither search strategy consistently achieves the highest classification accuracy.

\begin{figure}[t]
    \centering
    \includegraphics[width=\columnwidth]{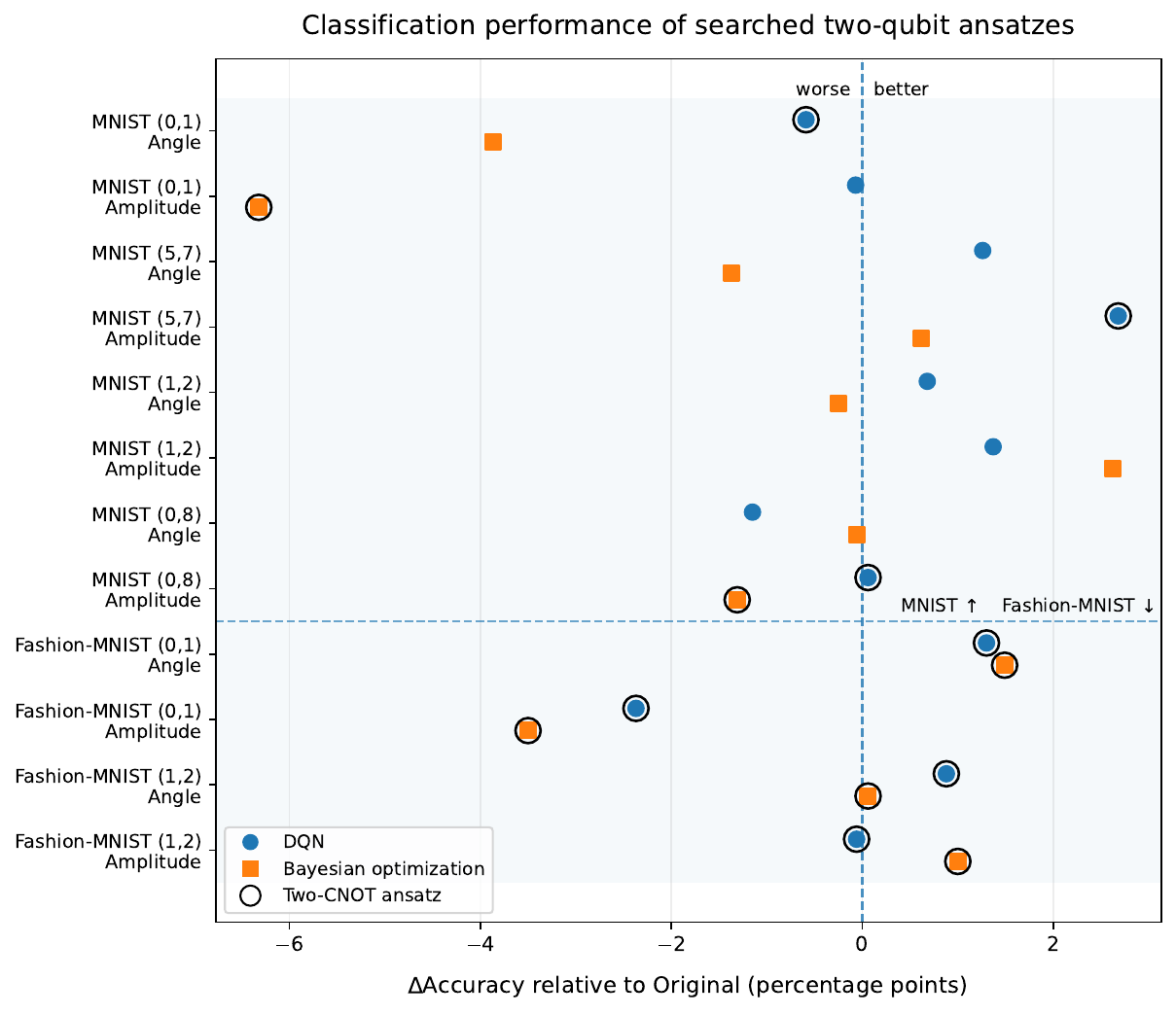}
    \caption{
    Classification accuracy differences between
    the two-qubit ansatzes identified by DQN
    and Bayesian optimization (BO) and the
    reference ansatz across different binary
    classification tasks and data-encoding
    strategies. The horizontal axis represents
    $\Delta\mathrm{Acc}
    =\mathrm{Acc}_{\mathrm{searched}}
    -\mathrm{Acc}_{\mathrm{reference}}$
    in percentage points (pp). Positive (negative) values indicate higher (lower) classification accuracy than the reference ansatz. Circles and squares denote architectures obtained by DQN and BO, respectively. An open circle surrounding a marker indicates that the corresponding searched ansatz contains only two CNOT gates.
    }
    \label{fig:qas_performance}
\end{figure}

Figure~\ref{fig:qas_performance}
summarizes the classification accuracy
differences relative to the reference
ansatz across the considered tasks
and encoding strategies.

The circuit structures of the selected
two-CNOT ansatzes are examined in the
following subsection.

\subsection{Compact ansatzes with two CNOT gates}
\label{subsec:compact_ansatz}

As indicated by the superscript $\star$ in
Table~\ref{tab:individual_results},
two-CNOT ansatzes are identified in
seven of the twelve task--encoding
settings using DQN and in six settings
using BO. Overall, thirteen of the
twenty-four selected ansatzes contain
only two CNOT gates, despite the absence
of an explicit circuit-complexity penalty
in the search objective.

\begin{figure}[t]
    \centering
    \includegraphics[width=0.95\columnwidth]
    {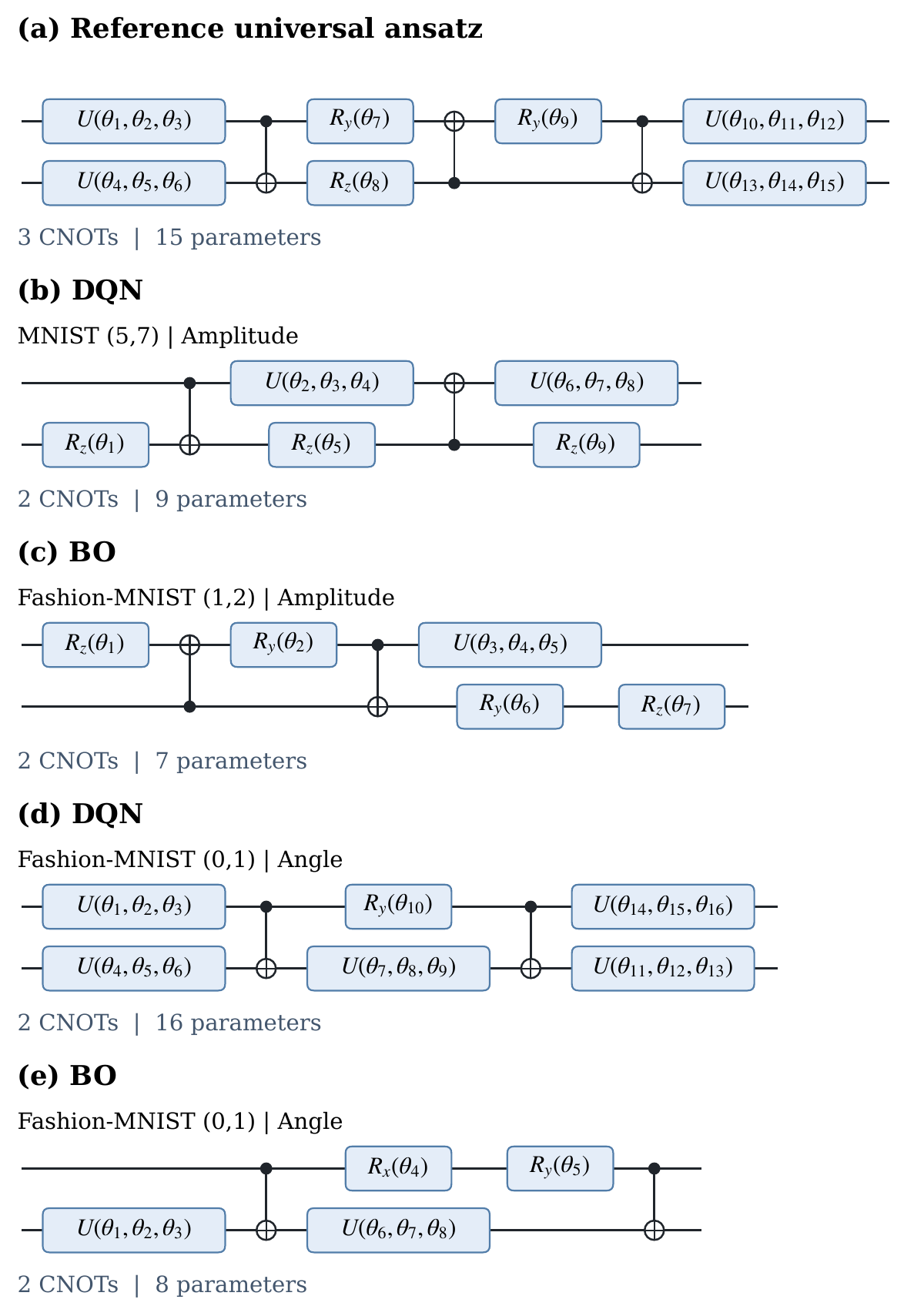}
    \caption{
    Comparison of the reference and selected
    two-qubit ansatzes.
    (a) Universal $SU(4)$ reference ansatz
    with three CNOT gates.
    (b)--(e) Representative two-qubit ansatzes
    with only two CNOT gates identified by
    DQN and Bayesian optimization across
    different classification tasks and
    data-encoding strategies.
    }
    \label{fig:searched_ansatzes}
\end{figure}

Figure~\ref{fig:searched_ansatzes} compares
the universal $SU(4)$ reference ansatz
with four representative two-CNOT ansatzes
identified across different search methods,
classification tasks, and encoding strategies.

Among the thirteen selected two-CNOT
ansatzes, seven achieve higher mean test
accuracies than the reference ansatz.
For MNIST $(0,8)$ with amplitude encoding,
the ansatz identified by DQN achieves
$94.56\%$, compared with $94.50\%$ for
the reference ansatz.
For Fashion-MNIST $(0,1)$ with angle encoding,
the two-CNOT architectures obtained by DQN
and BO achieve $91.93\%$ and $92.12\%$,
respectively, exceeding the reference
accuracy of $90.63\%$.

Competitive two-CNOT solutions are also
identified for Fashion-MNIST $(1,2)$.
Under angle encoding, the ansatz identified
by DQN achieves $97.13\%$, compared with
$96.25\%$ for the reference ansatz.
Under amplitude encoding, the ansatz obtained
using BO achieves $98.44\%$, exceeding the
reference result of $97.44\%$.
However, a compact gate structure does not
necessarily guarantee improved classification
performance. For example, on Fashion-MNIST
$(0,1)$ with amplitude encoding, the
two-CNOT ansatzes obtained by DQN and BO
achieve $95.38\%$ and $94.25\%$,
respectively, both below the reference
accuracy of $97.75\%$.

These results demonstrate that architecture
search can identify two-qubit ansatzes that
achieve higher classification accuracies than
the reference ansatz while requiring only
two CNOT gates. Some two-CNOT ansatzes
achieve slightly lower classification
accuracies than the reference ansatz.
Considering the potential benefits of
reducing entangling-gate counts on noisy
quantum devices, we argue that these
ansatzes may still represent competitive
alternatives to the universal two-qubit
reference ansatz.

\begin{table}[t]
\caption{
Mean test classification accuracies (\%) over five training runs with and without frozen first-layer transfer. L2--3 denotes the ansatz configuration of the second and third convolutional layers. Ref. and Trans. denote the reference and frozen source-trained first-layer configurations, respectively. Trans. L1 indicates the search method used to obtain the transferred ansatz. $\Delta_{\mathrm{reuse}}$ denotes the accuracy difference (Trans. minus Ref.) in percentage points (pp). A superscript $\star$ indicates a two-CNOT ansatz in the second and third convolutional layers, excluding the first layer.
}
\label{tab:transfer_comparison}
\setlength{\tabcolsep}{3pt}
\begin{ruledtabular}
\begin{tabular}{lllrrr}
L2--3 & Encoding & Trans. L1
& Ref. & Trans.
& reuse \\
\hline
\multicolumn{6}{c}{
\textbf{(a) MNIST (5,7)
 MNIST (1,2)}
} \\
\hline
DQN & Angle & DQN
& 93.44 & 94.37 & +0.93 \\
DQN & Amplitude & DQN
& 92.28 & 90.37 & -1.91 \\
BO & Angle & BO
& 94.69 & 93.69 & -1.00 \\
BO & Amplitude & BO
& 94.37 & 95.75 & +1.38 \\
Original & Angle & DQN
& 97.63 & 94.19 & -3.44 \\
Original & Amplitude & DQN
& 94.38 & 91.16 & -3.22 \\
\hline
\multicolumn{6}{c}{
\textbf{(b) Fashion-MNIST (0,1)
 MNIST (0,1)}
} \\
\hline
DQN & Angle & DQN
& 98.81 & 98.44 & -0.37 \\
DQN & Amplitude & DQN
& 97.88 & 96.69 & -1.19 \\
BO & Angle & BO
& 98.81 & 98.56 & -0.25 \\
BO & Amplitude & BO
& 98.50 & 96.25 & -2.25 \\
Original & Angle & DQN
& 98.24 & 98.62 & +0.38 \\
Original & Amplitude & DQN
& 99.32 & 98.30 & -1.02 \\
\end{tabular}
\end{ruledtabular}
\end{table}
\subsection{Cross-task reusability of frozen ansatzes}
\label{subsec:transfer_results}

We further investigate whether the
discovered two-qubit ansatzes can be
reused across different classification
tasks. Following the transfer protocol
described in Sec.~IV, both the architecture
and optimized variational parameters of
the source-trained first convolutional
layer are kept frozen during target-task
training.

We consider two transfer scenarios:
within-dataset transfer from MNIST $(5,7)$
to MNIST $(1,2)$, and cross-dataset
transfer from Fashion-MNIST $(0,1)$
to MNIST $(0,1)$.

For each scenario and encoding method,
we compare two first-layer configurations:
the reference ansatz (Ref.) and the frozen
source-trained ansatz (Trans.).
The transferred ansatzes are selected
from the source-task architectures
reported in Table~\ref{tab:individual_results}.

Three ansatz configurations are considered
for the second and third convolutional
layers: DQN, BO, and Original.
The transferred first-layer ansatz is
obtained using the corresponding search
method for the DQN and BO configurations,
and using DQN for the Original configuration.

In each paired comparison, the second
and third convolutional layers follow
the same architecture-search and
training protocols. However, the
resulting ansatz architectures may
differ between the reference and
transfer configurations.

To quantify the difference in classification
performance, we define
\begin{equation}
\Delta_{\mathrm{reuse}}
=
\mathrm{Acc}_{\mathrm{transfer}}
-
\mathrm{Acc}_{\mathrm{reference}},
\label{eq:reuse_accuracy}
\end{equation}
where $\mathrm{Acc}_{\mathrm{transfer}}$
and $\mathrm{Acc}_{\mathrm{reference}}$
denote the mean test accuracies on the
target task for the transfer and reference
configurations, respectively.
\begin{figure}[t]
    \centering
    \includegraphics[width=\columnwidth]
    {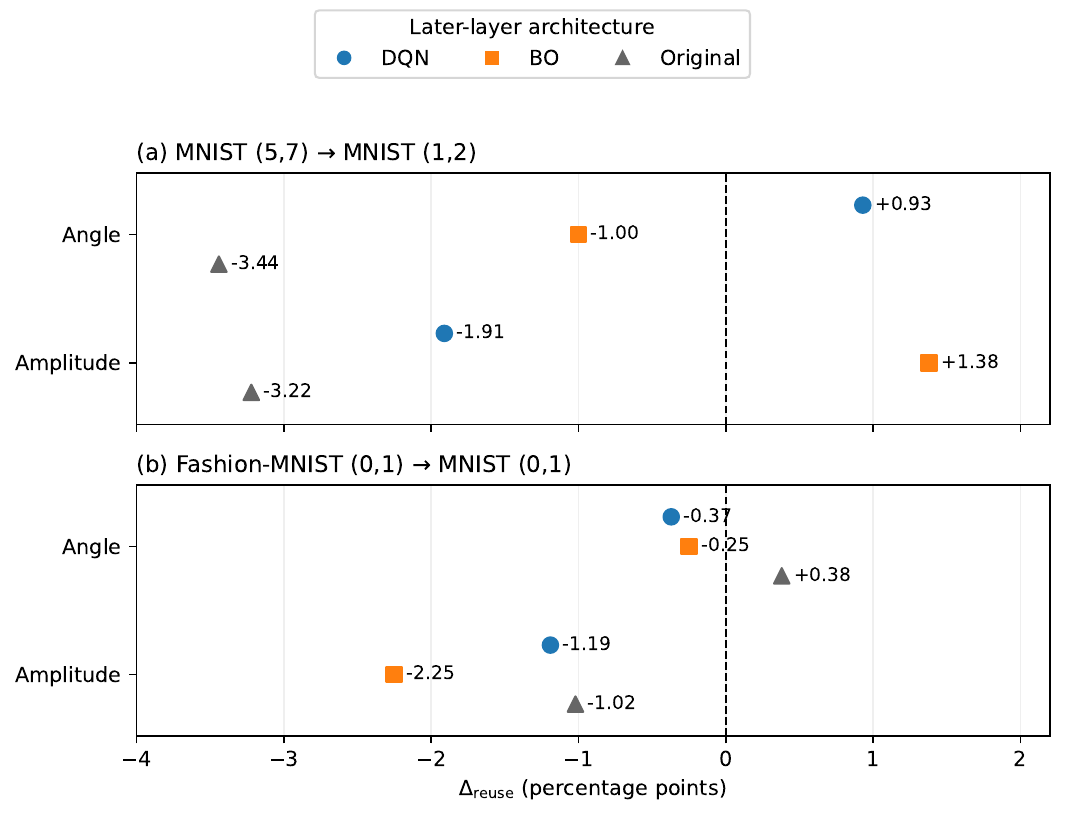}
    \caption{
    Mean test accuracy differences between
    reference and frozen first-layer transfer
    configurations:
    (a) within-dataset transfer from
    MNIST $(5,7)$ to MNIST $(1,2)$,
    and (b) cross-dataset transfer from
    Fashion-MNIST $(0,1)$ to MNIST $(0,1)$.
    The horizontal axis represents
    $\Delta_{\mathrm{reuse}}$ defined in
    Eq.~\eqref{eq:reuse_accuracy}, expressed
    in percentage points (pp).
    Circles, squares, and triangles denote
    the DQN, BO, and Original configurations
    of the second and third convolutional
    layers, respectively.
    Positive (negative) values indicate
    higher (lower) accuracy in the transfer
    configuration.
    }
    \label{fig:transfer_performance}
\end{figure}

Table~\ref{tab:transfer_comparison}
summarizes the mean test accuracies
for both transfer scenarios.
The corresponding accuracy differences
are illustrated in
Fig.~\ref{fig:transfer_performance}.

For within-dataset transfer, two of the
six configurations outperform their
corresponding reference configurations.
Under angle encoding, the DQN-based
configuration improves from $93.44\%$
to $94.37\%$. Under amplitude encoding,
the BO-based configuration improves
from $94.37\%$ to $95.75\%$.
The remaining four configurations
exhibit lower classification accuracies.

For cross-dataset transfer, the accuracy
differences range from $-2.25$ to
$+0.38$ percentage points.
Despite the decrease in accuracy in
five of the six configurations, all
transferred models retain mean test
accuracies above $96\%$.
The Original configuration under angle
encoding achieves $98.62\%$, compared
with $98.24\%$ for the reference
configuration.

Across the two transfer scenarios,
three of the twelve configurations
outperform their corresponding reference
configurations, whereas the remaining
nine exhibit lower accuracies.
These results suggest that the
reusability of frozen source-trained
ansatzes is configuration-dependent
and does not consistently lead to
improved classification performance.
Further investigation is required to
establish their transferability across
different classification tasks.

\section{Discussion and Conclusion}
\label{sec:discussion}

The results presented in Sec.~V demonstrate that
DQN- and BO-based quantum architecture search
can identify task-dependent two-qubit ansatzes
for QCNNs. Although the observed classification
performance of the searched ansatzes varies
across tasks and data-encoding strategies,
we argue that a fixed two-qubit ansatz may
not necessarily provide the most suitable
design for every application. In this context,
architecture search offers a systematic
alternative to applying a single predefined
gate sequence across different classification
problems.

The two search strategies explore the architecture space through different mechanisms:
DQN constructs candidate gate sequences sequentially by learning action values, whereas
BO proposes complete candidate architectures
using a probabilistic surrogate model. In our experiments, both search procedures are subject to the same wall-clock time limit, rather than a fixed number of architecture evaluations. Neither search strategy consistently outperforms the other or the reference ansatz across all considered tasks and encoding strategies.

A particularly interesting finding is the
identification of compact two-qubit ansatzes
containing only two CNOT gates. Despite their restricted representational capability compared with a universal $SU(4)$ ansatz, several of these circuits achieve competitive or improved classification accuracies. Importantly, neither the DQN reward nor the BO objective explicitly penalizes CNOT count or total gate count. These compact solutions therefore emerge as empirical outcomes of accuracy-driven architecture search rather than explicit circuit-complexity minimization. Evaluating the discovered ansatzes on real quantum hardware would provide
further insight into their practical advantages and limitations.
\begin{acknowledgments}
This work was supported by the National Science
and Technology Council (NSTC), Taiwan, under
Grant No.~NSTC 115-2112-M-033-009.
\end{acknowledgments}

\clearpage
\onecolumngrid

\appendix
\section{DDQN-based architecture-search algorithm}
\label{app:ddqn_algorithm}

Algorithm~\ref{alg:ddqn_search} summarizes
the DDQN-based quantum architecture-search
procedure described in Sec.~IV.
The algorithm incorporates sequential
ansatz construction, accuracy-based
experience replay, length-dependent
adaptive reward computation, and
DDQN network updates. The shortlisted architectures are subsequently
evaluated using the selection procedure
described in Sec.~\ref{subsec:ansatz_selection}.

\begin{algorithm}[!h]
\algorithmcaption{DDQN-based quantum architecture search}
\label{alg:ddqn_search}

\begin{algorithmic}[1]

\Require Gate set $\mathcal{G}$,
training dataset $\mathcal{D}_{\mathrm{tr}}$,
validation dataset $\mathcal{D}_{\mathrm{val}}$,
maximum sequence length $L_{\max}$,
maximum number of episodes $N_{\mathrm{ep}}$,
computational time limit $T_{\max}$,
shortlist size $M$,
DDQN hyperparameters

\Ensure Shortlisted architectures $\mathcal{S}_M$

\State Initialize policy network $Q(\cdot;\boldsymbol{w})$
\State Initialize target network $Q(\cdot;\boldsymbol{w}^{-})$
\State Initialize replay memory $\mathcal{M}$
\State Initialize accuracy baselines
       $\{C_{\mathrm{rec},n_a}\}_{n_a=1}^{L_{\max}}$
\State Initialize candidate archive
       $\mathcal{H}\gets\emptyset$
\State Initialize global search step $k\gets 0$

\For{$t=0,\ldots,N_{\mathrm{ep}}-1$}

    \State Initialize empty gate sequence $s$

    \For{$n_a=1,\ldots,L_{\max}$}

        \State Select gate $a\in\mathcal{G}$
               using the policy network

        \State Construct candidate ansatz
               $s'\gets s\oplus a$

        \State Train the candidate QCNN
               by minimizing the BCE loss

        \State Evaluate validation accuracy $c$

        \State Store $(s,a,c,s')$ in $\mathcal{M}$

        % \State Update $C_{\mathrm{rec},n_a}$
        \State Update $C_{\mathrm{rec},n_a}$
       using Eq.~\eqref{eq:recorded_accuracy_update}

        \State Sample a mini-batch $\mathcal{B}$
               from $\mathcal{M}$

        \State Determine $n_i=|s_i'|$
               for each sampled transition

        \State Recompute rewards $r_i^{(k)}$
               using Eq.~\eqref{eq:replay_reward}

        \State Compute DDQN targets using
               Eq.~\eqref{eq:ddqn_target}

        \State Optimize $\boldsymbol{w}$
               using the Huber loss in
               Eq.~\eqref{eq:dqn_loss}

        \State Update target-network parameters:
        \Statex \hspace{\algorithmicindent}
        $\boldsymbol{w}^{-}
        \gets
        \tau\boldsymbol{w}
        +(1-\tau)\boldsymbol{w}^{-}$

        \State Record $(s',c)$ in $\mathcal{H}$

        \State $s\gets s'$

        \State $k\gets k+1$

    \EndFor

    \If{computational time limit $T_{\max}$ is reached}
        \State \textbf{break}
    \EndIf

\EndFor

\State Select the top $M$ architectures
       from $\mathcal{H}$ according to
       their validation accuracies
\State \Return $\mathcal{S}_M$

\end{algorithmic}
\end{algorithm}

\clearpage
\twocolumngrid

\end{document}